\documentclass[twocolumn,prl,showpacs]{revtex4}
\usepackage[colorlinks,bookmarks=false,citecolor=blue,linkcolor=red,urlcolor=blue]{hyperref}
\usepackage{graphicx,epstopdf,color}
\usepackage{amsfonts}
\usepackage{amsmath,amssymb,mathrsfs}
\usepackage{bm}
\usepackage{ucs} % needed for utf8x
\usepackage[utf8x]{inputenc} 
\usepackage[T1]{fontenc}

\newcommand{\bs}[1]{\boldsymbol{#1}}
\newcommand{\braket}[2]{\left\langle #1 | #2 \right\rangle}

\newcommand{\bra}[1]{\left\langle#1\right|}
\newcommand{\ket}[1]{\left|#1\right\rangle}

\newcommand{\half}{$\frac{1}{2}$ }

\newcommand{\up}{\uparrow}
\newcommand{\dw}{\downarrow}
\newcommand{\vac}{\left|0\right\rangle}

\renewcommand{\a}{\alpha}
\renewcommand{\b}{\beta}
\renewcommand{\c}{\gamma}
\newcommand {\ea}{\eta_{\alpha}}
\newcommand {\eb}{\eta_{\beta}}
\newcommand {\ec}{\eta_{\gamma}}

\newcommand {\eab}{\bar\eta_{\alpha}}
\newcommand {\ebb}{\bar\eta_{\beta}}

\newcommand {\Sa}{S_{\alpha}}
\newcommand {\Sb}{S_{\beta}}

\newcommand {\bSa}{\bs{S}_{\alpha}}
\newcommand {\bSb}{\bs{S}_{\beta}}
\newcommand {\bSc}{\bs{S}_{\gamma}}

\def\s{\scriptscriptstyle}

\newlength{\ytlength}
\def\ie{{i.e.},\ }
\def\eg{{e.g.}\ }

\allowdisplaybreaks[1]
\begin{document}
%%%%%%%%%%%%% title, author, affiliation, abstract %%%%%%%%%%%%%%%%%
\title{Method to identify parent Hamiltonians for trial states}

\author{Martin Greiter}
\author{Vera Schnells}
\author{Ronny Thomale}
\affiliation{Institute for Theoretical Physics, University of
  W\"urzburg, Am Hubland, 97074 W\"urzburg, Germany}

\pagestyle{plain}

\begin{abstract}
  We describe a general method to identify exact, local parent
  Hamiltonians for trial states like quantum Hall or spin liquid
  states, which we have used extensively during the past decade.  It
  can be used to identify exact parent Hamiltonians, either directly
  or via the construction of simpler annihilation operators from which
  a parent Hamiltonian respecting all the required symmetries can be
  constructed.  Most remarkably, however, the method provides
  approximate parent Hamiltonians whenever an exact solution is not
  available within the space of presumed interaction terms.
\end{abstract}


%03.65.Vf Topological phases (quantum mechanics)
%74.20.-z Theories and models of superconducting state
%75.10.Jm Quantized spin models
%75.10.Pq Spin chain models
%75.10.Dg Spin Hamiltonians
% maybe FQHE or non-Abelian statistics??

\maketitle
%\section{Introduction}
\emph{Introduction.}---In the study of condensed matter systems with
conceptionally or even topologically non-trivial properties including
superconductors~\cite{Schrieffer64}, fractionally quantized Hall
fluids~\cite{laughlin83prl1395,haldane83prl605,Jain07}, or spin
liquids in
one~\cite{haldane88prl635,shastry88prl639,haldane-92prl2021,Greiter11}
or two
dimensions~\cite{kalmeyer-87prl2095,laughlin-90prb664,schroeter-07prl097202,greiter-09prl207203,greiter-14prb165125}
(1D or 2D), it has often been extremely helpful to resort to trial
wave functions which serve as paradigms for the %relevant
universality classes at hand.  These trial wave functions are usually
amenable to analytic formulations, and instruct us on the properties,
and in particular the quantum numbers, of the excitations above the
ground state.  Well known examples of such trial states are the BCS
wave function~\cite{Schrieffer64}, which supports Bogoliubov
quasiparticles, the Laughlin~\cite{laughlin83prl1395},
Moore--Read~\cite{moore-91npb362} and
Read--Rezayi~\cite{read-99prb8084} states in the quantum Hall effect,
which support fractionally charged quasiparticles with Abelian~\cite{arovas-84prl722} or
non-Abelian statistics~\cite{moore-91npb362,stern08ap204}, and the Gutzwiller ground
state~\cite{gutzwiller63prl159,metzner-87prl121,gebhardt-87prl1472} of
the Haldane---Shastry (HS)
model~\cite{haldane88prl635,shastry88prl639,haldane-92prl2021}, which
supports spinon excitations with half-fermi statistics.

In some cases, it is only possible to study these paradigms using
approximate parent Hamiltonians.  Whenever available, however, it is
highly desirable to construct exact parent Hamiltonians, and thus
elevate the paradigm from a wave function to an exact model.  This has
been accomplished for all the examples mentioned
above~\cite{richardson63pl277,richardson77jmp1802,haldane88prl635,shastry88prl639,haldane83prl605,greiter-92npb567,read-99prb8084,PhysRevX.5.041003},
and has been particularly rewarding in the case of the Gutzwiller
ground states of 1D spin chains, where the model turned out to be an
exact lattice realization of the SU(2)$_1$ Wess--Zumino--Witten
(WZW)
model~\cite{Witten84cmp455,affleck-87prb5291,%Fuchs92,
DiFrancescoMathieuSenechal97,GogolinNersesyanTsvelik98}.
This model was subsequently generalized from SU(2) to SU($M,N$)
supersymmetry~\cite{kuramoto-91prl1338,kawakami92prb3191},
% (where $N$ and $M$ labels the number of flavours or colors of charges
% and spins in the model),
and also to higher spin representations of SU(2)~\cite{Greiter11,nielsen-11jsmte11014,thomale-12prb195149},
 where the low energy sector is described by the SU(2)$_{k=2S}$ WZW
model.  
All these developments have been inspired not by the
Gutzwiller states directly, but by its parent Hamiltonian, which was
independently discovered by Haldane and Shastry~\cite{haldane88prl635,shastry88prl639}.

In this note, we describe a general, numerical method to obtain exact
parent Hamiltonians for given trial wave functions, which almost
trivially yields parent Hamiltonians for the Laughlin and for the
Gutzwiller wave functions discussed above.  The approach is in part similar to
recent proposals by Xi and Renard~\cite{arXiv:1712.01850} as well as by Chertkov
and Clark~\cite{arXiv:1802.01590}. Over the years, we have obtained
new results using this method for the hierachical quantum Hall
states~\cite{greiter94plb48} (with wave functions obtained either
through a composite fermion construction or through an explicit
condensation of quasiparticles in the hierachy), for the non-Abelian
chiral spin liquid
(NACSL)~\cite{ronnyphd,greiter-09prl207203,greiter-14prb165125}, and most
recently, for a new universality class of fractional topological
insulators~\cite{vera} we propose.  In the latter two examples, the
method not only revealed that there do not exist exact parent
Hamiltonians containing only the interaction terms we considered, but
provided us with meaningful approximate parent Hamiltonians, which
were instrumental to our studies.

\emph{General method.}---With these introductory remarks, we now turn
to the method itself.  Let $\ket{\psi_0}$ be a known trial ground
state for a finite system, of a system size amenable to exact
diagonalization studies.  We now wish to ask whether $\ket{\psi_0}$
is the exact ground state of a (local) model Hamiltonian specified by
a finite number $L$ of terms $H_i$ with unknown coefficients $a_i$,
\begin{align}
  \label{eq:h1}
  H=\sum_{i=1}^L a_i H_i,
\end{align}
and determine the coefficients.  To begin with, this requires that
$\ket{\psi_0}$ is an exact eigenstate,
\begin{align}
  \label{eq:s1}
  H \ket{\psi_0}=E_0\ket{\psi_0},
\end{align}
%or, with $H'=H+a_0$,
which we write as
\begin{align}
  \label{eq:s2}
%  H' \ket{\psi_0}=0.
  (H+a_0) \ket{\psi_0}=0.
\end{align}
Clearly, the additional variational parameter $a_0$ is to be
interpreted as $-E_0$.  Defining $H_0\equiv 1$, we may write this
compactly as
\begin{align}
  \label{eq:s3}
  \sum_{i=0}^L a_i H_i \ket{\psi_0}=0.
\end{align}
Since we are interested in identifying parent Hamiltonians for highly
correlated many body states, and the number of translationally
invariant $m$-body terms $H_i$ for a system with $N$ sites, scales roughly as $N^{m-1}$, the
dimension of the Hilbert space for system sizes with more than about 4
particles will in general be larger than the number of terms $L$.
This means that some special principle must be at work for each
solution of \eqref{eq:s3}.  In most applications, there is one or
several solutions due to conserved quantities (\eg total spin in a
spin system, total angular momentum for quantized Hall fluids on the
sphere), and an additional one if an exact parent Hamiltonian exists.

To find these solutions, we define the state vectors
$\ket{\psi_i}\equiv H_i\ket{\psi_0}$, and multiply \eqref{eq:s3} from
the left with the corresponding dual $\bra{\psi_j}$.  With
$M_{ji}\equiv \braket{\psi_j}{\psi_i}$, this yields
\begin{align}
  \label{eq:mji}
  \sum_{i=0}^L M_{ji}a_i=0\quad\text{for}\quad j=0,1,\ldots,L.
\end{align}
Obviously, there is one solution of \eqref{eq:mji} for each zero
eigenvalue of the $L+1$ dimensional, Hermitian matrix $M_{ji}$.
Substitution of the corresponding eigenvectors $a_i$ into
\eqref{eq:s3} yields operators annihilating the ground state,
which enable us to extract the desired parent Hamiltonian
\eqref{eq:h1}.
% 
% Similar approaches has
% recently been suggested independently by Xi and
% Renard~\cite{arXiv:1712.01850} by Chertkov and
% Clark~\cite{arXiv:1802.01590}.  
% 
Even though this may come across as a trivial observation, the models
one can obtain with this method are in general highly non-trivial.

Possibly the most outstanding feature is that, according to our long
standing experience, the method usually yields a highly non-trivial
approximate parent Hamiltonian if no exact one exists within the
operator space spanned by the $H_i$'s.  In these cases, there are
likewise one or several zero eigenvalues due to conserved quantities,
and one small or very small, nonzero eigenvalue.  The eigenstate
corresponding to this eigenvalue defines the approximate Hamiltonian.

An obvious drawback is that the method guarantees that $\ket{\psi_0}$
is an exact or approximate eigenstate of $H$, but not that it is the
ground state.  This has hence to be verified {\it a posteriori} by exact, numerical
diagonalization of $H$.  Our experience here is that whenever an exact
parent Hamiltonian exists, it will have $\ket{\psi_0}$ as its unique
ground state.  In the case of approximate solutions, we have sometimes
encountered situations where the method suggested operators for which
$\ket{\psi_0}$ has only been an approximate eigenstate, not the ground
state.  In the cases we have studied, however, it was always possible
to find a suitable set of operators $H_i$ such that the method
converged on an approximate parent Hamiltonian for the ground state.

\emph{Example: The Haldane--Shastry Hamiltonian.}---The ground state
of the model can be obtained by Gutzwiller
projection from a completely filled one-dimensional band, which in
total contains as many spin \half fermions as there are lattice
sites~\cite{gutzwiller63prl159,%kaplan-82prl889,gros-87prb381,
metzner-87prl121,gebhardt-87prl1472}:
\begin{equation}
  \label{eq:hsgw}
  \ket{\psi^{\s\text{HS}}_{0}}
  =\text{P}_{\s\text{GW}}\ket{\psi^{\s N}_{\s\text{SD}}},\quad
  \ket{\psi^{\s N}_{\s\text{SD}}}\equiv 
  \prod_{q\in\mathcal{I}} c_{q\up}^\dagger c_{q\dw}^\dagger\ket{0},
\end{equation}
where the interval $\mathcal{I}$ contains $M=\frac{N}{2}$ adjacent
momenta, and the Gutzwiller projector 
$P_{\s\text{GW}}$ eliminates doubly occupied sites.
% \begin{equation}
%   \label{eq:hsgwp}
%   \text{P}_{\s\textrm{GW}} \equiv \prod_{i=1}^N
%   \big(1-c^\dagger_{i\up}c^{\phantom{\dagger}}_{i\up}
%   c^\dagger_{i\dw}c_{i\dw} \big)
% \end{equation}
% eliminates configurations with more than one particle on any site.

While it is irrelevant to the applicability of the numerically
executed method proposed above, a different formulation of the
Gutzwiller ground state is convenient for the discussion below.
Consider a spin \half chain with periodic boundary conditions and an
even number of sites $N$ on a unit circle embedded in the complex
plane:
%\begin{equation}
%\begin{picture}(280,70)(-20,-35)
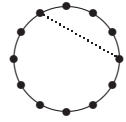
\begin{center}
\begin{picture}(320,50)(-38,-30)
\put(0,0){\circle{100}}
%\put(20,0){\circle*{3}}
\put( 20.0,   .0){\circle*{3}}
\put( 17.3, 10.0){\circle*{3}}
\put( 10.0, 17.3){\circle*{3}}
\put(   .0, 20.0){\circle*{3}}
\put(-10.0, 17.3){\circle*{3}}
\put(-17.3, 10.0){\circle*{3}}
\put(-20.0,   .0){\circle*{3}}
\put(-17.3,-10.0){\circle*{3}}
\put(-10.0,-17.3){\circle*{3}}
\put(   .0,-20.0){\circle*{3}}
\put( 10.0,-17.3){\circle*{3}}
\put( 17.3,-10.0){\circle*{3}}
\qbezier[20]( 20.0,   .0)(  5.0,8.65)(-10.0, 17.3)
\put(50,12){\makebox(0,0)[l]
{$N$\ sites with spin \half on unit circle: 
}}
%$\displaystyle \eta_\alpha=e^{i\frac{2\pi}{N}\alpha }$}}
\put(50,-12){\makebox(0,0)[l]
{$\displaystyle \eta_\alpha=e^{\text{i}\frac{2\pi}{N}\alpha }$
\ \ with\ $\alpha = 1,\ldots ,N$}}
\end{picture}
\end{center}
%\end{equation}
The ground state \eqref{eq:hsgw} can be written as
\begin{equation}
  \label{eq:hsket}
  \ket{\psi^{\s\text{HS}}_{0}}\hspace{0pt}=\hspace{-4pt}
  \sum_{\{z_1,\ldots ,z_M\}}\hspace{-4pt}\psi^{\s\text{HS}}_{0} 
  (z_1,\ldots ,z_M)\,{S}^+_{z_1}\ldots {S}^+_{z_M} 
  |\underbrace{\dw\dw\ldots\dw}_{N \text{\ spins\ } \dw}
  \rangle,
\end{equation}
where the sum extends over all possible ways to distribute the 
$M=\frac{N}{2}$ $\up$-spin coordinates $z_i$ on the unit circle and
\begin{equation}
  \label{eq:hspsi0}
  \psi^{\s\text{HS}}_{0}(z_1,\ldots ,z_M) = 
  \prod_{i<i}^M\,(z_i-z_j)^2\,\prod_{i=1}^M\,z_i\,. 
\end{equation}
We now search numerically for an exact parent Hamiltonian which is
invariant under all the symmetries of the ground state, \ie
translations, SU(2) spin rotations, time reversal (T), and parity (P).
It is further reasonable to first try an Ansatz with two-body
interactions only.  (In fact, the only SU(2) invariant three spin
interaction term for spin one-half %$S=\frac{1}{2}$
is $\text{i}\bs{S}_{\alpha} (\bs{S}_{\beta}\times\bs{S}_{\gamma})$,
which violates T.)  Following the notation in
\eqref{eq:h1}, we write
\begin{align}
  \label{eq:hsham1}
  H=\sum_{i=1}^{N/2} a_i H_i,\quad 
  H_i=\sum_{\alpha=1}^N {\boldsymbol{S}}_\alpha {\boldsymbol{S}}_{\alpha+i}.
\end{align}
Numerical execution of the steps described above for a chain with $N\ge 8$
sites yields two zero eigenvalues of the matrix $M_{ji}$ of
\eqref{eq:mji}.  The corresponding eigenvectors yield, upon rewriting
in a more convenient form, the ground state annihilation operators
\begin{align}
  \label{eq:hsham1}
  H^{\text{a}}={\boldsymbol{S}_{\text{tot}}^2},\quad %\text{with}\quad
  {\boldsymbol{S}_{\text{tot}}}\equiv \sum_{\alpha=1}^N {\boldsymbol{S}}_\alpha,
\end{align}
and
\begin{align}
  \label{eq:hsham}
  H^{\text{b}}=%{H}^{\s\text{HS}}\hspace{-2pt}-\hspace{-2pt}E_0,\
%  {H}^{\s\text{HS}}= 
  -E_0+
  \left(\frac{2\pi}{N}\right)^{\hspace{-2pt}2}
  \sum^N_{\alpha <\beta}\,
  \frac{{\boldsymbol{S}}_\alpha {\boldsymbol{S}}_\beta 
  }{\left|\eta_\alpha-\eta_\beta \right|^2}\,,
\end{align}
where $\left|\eta_\alpha-\eta_\beta \right|$ is the chord distance
between the sites $\alpha$ and $\beta$, and
$E_0=-\frac{\pi^2}{24}\left(N+\frac{5}{N}\right)$.  While
$H^{\text{a}}$ just confirms that the ground state is a spin singlet,
$H^{\text{b}}$ is the model Hamiltonian discovered by Haldane and
Shastry~\cite{haldane88prl635,shastry88prl639}.

\emph{First generalization: Ground state annihilation
  operators.}---While the method in its most direct form works
extremely well for 1D models (such as spin chains or 2D electrons
confined to a Landau level) with two-body interactions, and is still
feasible for 1D models with three-body interactions, it is less so for
higher dimensions.  In some instances it has been extremely helpful to
employ the method to identify not the coefficients in a model
Hamiltonian directly, but in an annihilation operator for the ground
state.  Such an operator can be much simpler than the Hamiltonian, is
not required to share any of the symmetries of the ground state,
and does not need to be Hermitian.  Once the operator is known, it is
usually possible to construct a local and positive semi-definite
parent Hamiltonian from it.  Returning to our example of the HS model,
an operator of this kind is
\begin{align}
  \label{eq:ann1}
  \Omega_\alpha =\sum_{i=1}^{N-1} 
  a_{\alpha,i}\, H _{\alpha,i},\quad  H_{\alpha,i} = S_\alpha^- S^-_{\alpha+i}.
\end{align}
Note that even though the operators $H_{\alpha,i}$ are no longer
Hermitian, the matrices $M_{\alpha, ji}$ of \eqref{eq:mji} still are.
We now find three zero eigenvalues for each $\alpha$, which yield the
ground state annihilations operators
\begin{align}
  \label{eq:anna}
  \Omega_\alpha^{\text{a}} = %\sum_{i=1}^{N-1} S_\alpha^- S^-_{\alpha+i} 
  S_\alpha^- S _{\text{tot}}^-,
\end{align}
where we have used $(S_\alpha^-)^2=0$, 
\begin{align}
  \label{eq:annb}
  \Omega_\alpha^{\text{b}} = \sum_{\substack{\beta=1\\[2pt]\beta\ne\alpha}}^N 
  \frac{1}{\ea-\eb} \Sa^- \Sb^-,\quad\text{and}\quad 
  \Omega_\alpha^{\text{c}} = \bigl(\Omega_\alpha^{\text{b}}\bigr)^*.
\end{align}
It is then an elementary exercise~\cite{Greiter11} to show that the T and P invariant
scalar component (scalar with regard to SU(2) spin rotations) of the
translationally invariant, Hermitian and semi-definite positive
operator
\begin{align}
  \label{eq:hsbefore}
  H_{\text{intermediate}}
  =\sum_{\alpha=1}^N {\Omega_\alpha^{\text{b}}}^\dagger {\Omega_\alpha^{\text{b}}}
\end{align}
is, up to an overall normalization, equal to \eqref{eq:hsham}.  Since
$\Omega_\alpha^{\text{c}}$ is just the T or P conjugate of
$\Omega_\alpha^{\text{b}}$, it yields the same parent Hamiltonian.

The advantages of an approach via an annihilation operator of the kind
\eqref{eq:ann1} over the direct approach \eqref{eq:hsham1} become
apparent as we consider models which are not as readily obtained as
the HS model \eqref{eq:hsham}.  
% 
% For example, for the
% higher spin $S$ generalizations of the Gutzwiller state, we can
% numerically determine the coefficients $a_i$ in \eqref{eq:ann1} with
% $H _{\alpha,i} = (S_\alpha^-)^{2S} S^-_{\alpha+i}$, which then turn
% out to be equivalent to those found in \eqref{eq:anna} and
% \eqref{eq:annb} for spin one half.  
% 
Consider the higher spin $S$ generalizations~\cite{greiter02jltp1029}
of the Gutzwiller state,
\begin{align}
  \label{eq:psi0schwingerS}
  \ket{\psi^{S}_0}
  =\Big(\Psi^{\s\text{HS}}_0\big[a^\dagger ,b^\dagger\big]\Big)^{2S}\vac,
\end{align} 
where $\Psi^{\s\text{HS}}_0\big[a^\dagger ,b^\dagger\big]$ is the
operator generating the HS ground state in terms of Schwinger bosons,
such that 
\begin{align}
  \label{eq:psi0schwinger}
  \ket{\psi^{\s\text{HS}}_0}
  =\Psi^{\s\text{HS}}_0\big[a^\dagger ,b^\dagger\big]\vac.
\end{align} 
If we view the HS ground state \eqref{eq:hspsi0} as a bosonic
Laughlin state for spin-flip operators, we would view 
\eqref{eq:psi0schwingerS} as a bosonic Read---Rezayi
state for renormalized spin-flips.
For \eqref{eq:psi0schwingerS}, we can numerically determine the
coefficients $a_i$ in \eqref{eq:ann1} with
$H _{\alpha,i} = (S_\alpha^-)^{2S} S^-_{\alpha+i}$, which then turn
out to be equivalent to those found in \eqref{eq:anna} and
\eqref{eq:annb} for spin~$\frac{1}{2}$.  Following the same steps as above,
this yields the parent Hamiltonian~\cite{Greiter11}
\begin{align}
  \label{eq:spinSham}
  H^{S} =\frac{2\pi^2}{N^2}
  \Bigg[
  &\sum_{\substack{\a\ne\b}}^N
    \frac{\bSa\bSb}{\vert\ea-\eb\vert^2}
    -\frac{1}{2(S+1)(2S+3)}
  \nonumber\\[0.2\baselineskip] 
  &  \sum_{\substack{\a,\b,\c\\ \a\ne\b,\c}}^N
  \frac{(\bSa\bSb)(\bSa\bSc) + (\bSa\bSc)(\bSa\bSb)}{(\eab-\ebb)(\ea-\ec)}
  \Bigg]
\end{align}
with ground state energy
\begin{align}
  \label{eq:spinSenergy}
  E_0^{S} &=-\frac{2\pi^2}{N^2}\frac{S(S+1)^2}{2S+3}\,\frac{N (N^2+5)}{12}.
\end{align}
Obviously, it would be much more difficult to obtain the coefficients
in \eqref{eq:spinSham} directly with our numerical method than it is
with annihilation operators.  The direct method, however, does convey
the information that an exact parent Hamiltonian with three-body
terms of the form in \eqref{eq:spinSham} does exist for the higher
spin states.

% \begin{align}
%   \label{eq:i:a:Omegadef}
%   \OaHS =\sum_{\substack{\beta=1\\[2pt]\beta\ne\alpha}}^N 
%   \frac{1}{\ea-\eb} \Sa^- \Sb^-,\qquad 
%   \OaHS \ket{\psi^{\s\text{HS}}_{0}} = 0 \quad\forall\, \alpha.
% \end{align}

\emph{Second generalization:  
Approximate parent Hamiltonians.}---Possibly the most important
feature of our method is that it delivers approximate parent
Hamiltonians whenever an exact parent Hamiltonian for the trial
ground state is not available in the space spanned by the terms
$H_i$ one considers.  More often than not, this situation arises
because no simple, local, analytically amenable parent Hamiltonian 
exists for the state in question.  Examples for such a situation are
provided by the hierachy wave functions of the quantized Hall effect,
which is also the instance where one of us applied this method
first~\cite{greiter94plb48}, or for the NACSL~\cite{greiter-09prl207203}.

As explained in the context of the general method above, in situations
where no exact, but an approximate, parent Hamiltonian can be
constructed with the terms $H_i$ included in \eqref{eq:h1}, the
eigenvector associated with the smallest non-zero eigenvalue of
$M_{ji}$ usually provides such an approximate Hamiltonian $H$.  The
result, however, will slightly depend on the relative
normalizations $w_i$ of the operators $H_i$ used in the numerical
procedure.
% We refer to these normalizations as precondition weights $w_i$.
In this context, however, the optimal solution will depend on what one
desires to optimize.  This could be the relative variance of the
ground state energy
\begin{align}
  \label{eq:var}
  \frac{\bra{\psi_0}H^2\ket{\psi_0}-\bra{\psi_0}H\ket{\psi_0}^2}
  {\bra{\psi_0}H\ket{\psi_0}^2},
\end{align}
the overlap $\braket{\psi}{\psi_0}$ between the exact ground state
$\ket{\psi}$ of $H$ and the reference trial state $\ket{\psi_0}$, or
the similarity between the correlators $h_i=\bra{\psi}H_i\ket{\psi}$
and $h_{0,i}=\bra{\psi_0}H_i\ket{\psi_0}$.  (When we applied the method
to the NACSL~\cite{greiter-09prl207203}, our point was to show that
we can find a local, approximate Hamiltonian with a gap between the
three (in the TD limit topologically degenerate) ground states and
the remaining spectrum.  The size of this gap was hence 
a parameter we considered as well.)

In most applications we studied, the most naive application of the
method designed for the identification of an exact parent Hamiltonian
provided us with remarkably accurate approximations whenever no exact
solutions were available.  If one then desires to optimize the
Hamiltonian specified by the set of parameters
$[a_i]\equiv (a_0,a_1,\ldots ,a_L)$ further, one may apply a Newton
scheme, as follows.
We illustrate the method here for an optimization of the similarity in
the correlators, as this usually optimizes variance and overlap as
well.  To begin with, we choose a set of weights $[w_i]$, and
another set $[w_i']$, where only a single weight $w_j$ differs by a
small parameter $\delta_j$.  We then evaluate the corresponding
coefficients $[a_i]$ and $[a_i']$, and from there $[h_i]$ and
$[h_i']$.  This yields the $j$-th row of the derivative matrix
\begin{align}
  \frac{\partial h_i}{\partial w_j}\equiv \frac{h_i'-h_i}{\delta_j}.
  \nonumber  
\end{align}
As a next step, we solve the linear equation 
\begin{align}
  \sum_{j=0}^L \frac{\partial h_i}{\partial w_j} \Delta w_j =h_{0,i}-h_i
  \nonumber  
\end{align}
for the shifts $\Delta w_j$ we would require if we assume a linear
dependence.  The procedure can then be repeated with the adjusted
weights $[w_i +\Delta w_i]$ until it has converged.  In the examples
we considered, however, a single iteration was sufficient.  Whenever
adjustments of the weights $[w_i]$ are insufficient to induce the
desired changes in the correlators, one possible route is to follow
the same steps with infinitesimal variations in the coefficients
$[a_i]$.  Usually, one needs to adjust nuances of the method to the
problem one is considering.  For example, it is sometimes better to
include $w_0$ and $a_0$ in the optimization, while in other situations
it is better to take $a_0$ constant, if not zero to start with.
% (Sometimes we have to set $a_0=0$ from the beginning.)
We have also encountered examples where the optimization worked better
when we adjusted the weights not on a linear, but on a logarithmic
scale, a change which is fully implemented by taking $e^{w_i}H_i$
instead of $w_iH_i$ for the re-normalized operators in
$H=\sum_i a_i w_iH_i$ and $\ket{\psi_i}=w_iH_i\ket{\psi_0}$.  The
procedure we have outlined here hence should be taken mostly as an
inspiration to find an adequate algorithm for the problem one is
interested in.

The approximate method we just outlined is heuristic and crude, but
has been highly successful in our experience.  The reader might ask at
this point whether a more scholarly approach does not offer itself.
One possible avenue we have explored is to minimize the variance
\eqref{eq:var} by maximizing $\bra{\psi_0}H\ket{\psi_0}^2$ subject to
the constraints $\braket{\psi_0}{\psi_0}=\bra{\psi_0}H^2\ket{\psi_0}=1$
with $H$ given by \eqref{eq:h1}.  This yields
\begin{align}
\label{eq:scholarly}
  \sum_{i=1}^L M_{ji}a_i = -M_{j0} a_0\quad \text{for}\quad j=1,\ldots,L, 
\end{align}
where $a_0$ is now a normalization constant given by
\begin{align}
\label{eq:scholarlynorm}
  a_0^{-2}=\sum_{i,j=1}^L M_{0i}(M^{-1})_{ij}M_{j0}. 
\end{align}
Note that since $a_0$ only affects the overall normalization of the
parent Hamiltonian, we do not need to evaluate
\eqref{eq:scholarlynorm} in practical applications.  Instead, we may
set $a_0=1$ in \eqref{eq:scholarly}.  In some of the examples we have
investigated, the Hamiltonian corresponding
to the solution of \eqref{eq:scholarly} for $a_i$ was more accurate
than the one obtained with the previous method, \ie via the lowest
non-zero eigenvalue of \eqref{eq:mji}.  In general, however, this
method has not been as stable and robust as the previous one.

\emph{Conclusion.}---We have introduced a method to identify where
available exact, but in general approximate, parent Hamiltonians for
known trial wave functions.  It is particularly useful when the trial
states describe paradigms of fractionally quantized or topologically
ordered, many-particle states.  In the examples we studied, the most
naive application of the method provided us already with compelling
approximative Hamiltonians whenever exact Hamiltonians did not exist
within the space of the interaction terms we considered.  Since the
effort required to optimize these approximations is very manageable,
we explained and illustrated one optimization procedure in detail.  Different physical
problems usually require different approximations, and the
procedure we outline is not universally applicable. We do believe,
however, that the method in general will be of vital use in many
different areas of physics that concern themselves with trial states,
and yet unknown microscopic models associated with them.

%\vfill\eject %\phantom{r}\vfill\eject\phantom{r}\vfill\eject\vfill %\phantom{p}\eject\vfill 

\begin{acknowledgments}
  This work was supported by the European Research Council (ERC) starting grant 
  TOPOLECTRICS under ERC-StG-Thomale-336012. 
\end{acknowledgments}

%\appendix* 
%\section{Transformation properties of $G_{\be,\pbe}[z,\pz]$}

%\vfill\eject\vfill %\phantom{p}\eject\vfill

%\bibliographystyle{../../bib/prsty}\bibliography{../../bib/book,../../bib/paper,../../bib/proc,../../bib/unpub,../../bib/htc}

%\bibliographystyle{../../bib/prsty}\bibliography{../../bib/book,../../bib/paper,../../bib/proc,../../bib/unpub,../../bib/htc,add-refs}

\begin{thebibliography}{10}

\bibitem{Schrieffer64}
J.~R. Schrieffer, {\em Theory of {S}uperconductivity} (Benjamin/Addison Wesley,
  New York, 1964).

\bibitem{laughlin83prl1395}
R.~B. Laughlin, Phys. Rev. Lett. {\bf 50},  1395  (1983).

\bibitem{haldane83prl605}
F.~D.~M. Haldane, Phys. Rev. Lett. {\bf 51},  605  (1983).

\bibitem{Jain07}
J. Jain, {\em Composite Fermions} (Cambridge University Press, Cambridge,
  2007).

\bibitem{haldane88prl635}
F.~D.~M. Haldane, Phys. Rev. Lett. {\bf 60},  635  (1988).

\bibitem{shastry88prl639}
B.~S. Shastry, Phys. Rev. Lett. {\bf 60},  639  (1988).

\bibitem{haldane-92prl2021}
F.~D.~M. Haldane, Z.~N.~C. Ha, J.~C. Talstra, D. Bernard, and V. Pasquier,
  Phys. Rev. Lett. {\bf 69},  2021  (1992).

\bibitem{Greiter11}
M. Greiter, {\em Mapping of Parent {H}amiltonians}, Vol.~244 of {\em Springer
  Tracts in Modern Physics} (Springer, Berlin/Heidelberg, 2011),
  {a}rXiv:1109.6104.

\bibitem{kalmeyer-87prl2095}
V. Kalmeyer and R.~B. Laughlin, Phys. Rev. Lett. {\bf 59},  2095  (1987).

\bibitem{laughlin-90prb664}
R.~B. Laughlin and Z. Zou, Phys. Rev. B {\bf 41},  664  (1990).

\bibitem{schroeter-07prl097202}
D.~F. Schroeter, E. Kapit, R. Thomale, and M. Greiter, Phys. Rev. Lett. {\bf
  99},  097202  (2007).

\bibitem{greiter-09prl207203}
M. Greiter and R. Thomale, Phys. Rev. Lett. {\bf 102},  207203  (2009).

\bibitem{greiter-14prb165125}
M. Greiter, D.~F. Schroeter, and R. Thomale, Phys. Rev. B {\bf 89},  165125
  (2014).

\bibitem{moore-91npb362}
G. Moore and N. Read, Nucl. Phys. B {\bf 360},  362  (1991).

\bibitem{read-99prb8084}
N. Read and E. Rezayi, Phys. Rev. B {\bf 59},  8084  (1999).

\bibitem{arovas-84prl722}
D. Arovas, J.~R. Schrieffer, and F. Wilczek, Phys. Rev. Lett. {\bf 53},  722
  (1984).

\bibitem{stern08ap204}
A. Stern, Annals of Physics {\bf 323},  204   (2008).

\bibitem{gutzwiller63prl159}
M.~C. Gutzwiller, Phys. Rev. Lett. {\bf 10},  159  (1963).

\bibitem{metzner-87prl121}
W. Metzner and D. Vollhardt, Phys. Rev. Lett. {\bf 59},  121  (1987).

\bibitem{gebhardt-87prl1472}
F. Gebhard and D. Vollhardt, Phys. Rev. Lett. {\bf 59},  1472  (1987).

\bibitem{richardson63pl277}
R. Richardson, Phys. Lett. {\bf 3},  277  (1963).

\bibitem{richardson77jmp1802}
R. Richardson, J. Math. Phys {\bf 18},  1802  (1977).

\bibitem{greiter-92npb567}
M. Greiter, X.~G. Wen, and F. Wilczek, Nucl. Phys. B {\bf 374},  567  (1992).

\bibitem{PhysRevX.5.041003}
C.~H. Lee, Z. Papi\ifmmode~\acute{c}\else \'{c}\fi{}, and R. Thomale, Phys.
  Rev. X {\bf 5},  041003  (2015).

\bibitem{Witten84cmp455}
E. Witten, Commun. Math. Phys. {\bf 92},  455  (1984).

\bibitem{affleck-87prb5291}
I. Affleck and F.~D.~M. Haldane, Phys. Rev. B {\bf 36},  5291  (1987).

\bibitem{DiFrancescoMathieuSenechal97}
P. Di~Francesco, P. Mathieu, and D. S\'{e}n\'{e}chal, {\em Conformal Field
  Theory} (Springer, New York, 1997).

\bibitem{GogolinNersesyanTsvelik98}
A.~O. Gogolin, A.~A. Nersesyan, and A.~M. Tsvelik, {\em Bosonization and
  Strongly Correlated Systems} (Cambridge University Press, Cambridge, 1998).

\bibitem{kuramoto-91prl1338}
Y. Kuramoto and H. Yokoyama, Phys. Rev. Lett. {\bf 67},  1338  (1991).

\bibitem{kawakami92prb3191}
N. Kawakami, Phys. Rev.~B {\bf 46},  3191  (1992).

\bibitem{nielsen-11jsmte11014}
A.~E.~B. Nielsen, J.~I. Cirac, and G. Sierra, J. Stat. Mech.: Theory and
  Experiment {\bf 11},  P11014  (2011).

\bibitem{thomale-12prb195149}
R. Thomale, S. Rachel, P. Schmitteckert, and M. Greiter, Phys. Rev. B {\bf 85},
   195149  (2012).

\bibitem{arXiv:1712.01850}
X. Qi and D. Ranard, arXiv:1712.01850.

\bibitem{arXiv:1802.01590}
E. Chertkov and B.~K. Clark, arXiv:1802.01590.

\bibitem{greiter94plb48}
M. Greiter, Phys. Lett. B {\bf 336},  48  (1994).

\bibitem{ronnyphd}
R. Thomale, Ph.D. thesis, Karlsruhe, 2008.

\bibitem{vera}
V. Schnells, R. Thomale, and M. Greiter, manuscript in preparation.

\bibitem{greiter02jltp1029}
M. Greiter, J.~Low Temp. Phys. {\bf 126},  1029  (2002).

\end{thebibliography}

%\vfill\eject
\end{document}